\documentclass[prl,twocolumn,showpacs]{revtex4}
\usepackage{graphicx,amsmath,amssymb}

\usepackage{graphicx,amsmath,amssymb}

\begin{document}

\title{Nonconservative forcing and diffusion in refractive optical traps}

\author{Ingmar Saberi$^\dagger$ and Fred Gittes$^{*}$}
\address{Department of Physics \& Astronomy, Washington State University
Pullman, WA 99164-2814, USA. $^\dagger$Present address:
Department of Physics, California Institute of Technology, 
Pasadena, CA 91125, USA.
$^*$Corresponding author: gittes@wsu.edu
}


\begin{abstract}
Refractive optical trapping forces can be nonconservative in
the vicinity of a stable equilibrium point even in the absence of
radiation pressure.  We discuss how nonconservative 3D force
fields, in the vicinity of an equilibrium point, reduce to
circular forcing in a plane; a simple model of such forcing is
the refractive trapping of a sphere by a four rays.  We discuss
in general the diffusion of an anisotropically trapped,
circularly forced particle and obtain its spectrum of motion.
Equipartition of potential energy holds even though the
nonconservative flow does not follow equipotentials of the trap.
We find that the dissipated nonconservative power is proportional
to temperature, providing a mechanism for a runaway heating
instability in traps.
\end{abstract}




\maketitle

\section{Introduction}

Nonconservative fields of optical force on optically trapped
particles have long been predicted to occur \cite{Ashkin}.  These
force fields are locally nonpotential in character so that net
work is done on a particle even in microscopic closed paths.
This differs from, say, electromotive force around a circuit
where only the global potential is not definable\cite{Panofsky}.
In a locally nonconservative force field, external power is
continuously coupled to particle motion, leading to dissipation
and heating even when the particle is localized about a stable
point of the force field.

Radiation pressure on trapped particles can be one source of nonconservative
forcing.  In recent experiments \cite{Roichman,Sun} nonconservative toroidal
circulation of optically trapped particles caused by axial radiation pressure
was observed and theoretically analyzed.  Recent theoretical work \cite{Simpson}
has shown how, for nonspherical objects, nonconservative motion arises in
coordinates of angle and translation.  

Here we show that a simple spatial 3D nonconservative force field, 
circular rather than toroidal, occurs even in the absence of radiation
pressure. As a physical model leading to this force,
we describe a sphere trapped by refracting rays.
Within any such circular-forcing model, and generalizing to an anisotropic
trapping force, we derive the spectrum of motion and the thermal signatures
of nonconservative circular forcing.

\section{Geometry of nonconservative forces}

Before narrowing the discussion to optical trapping, we ask what simple generic
statements can be made about nonconservative forces.  Powerful classification
schemes are available from differential geometry
\cite{Edelen,Burke,Frankel} but to use these one must correctly identify
forces as fields, not of vectors, but of differential 1-forms.  This means
that force components such as $f_x$, $f_y$, and $f_z$ properly take their
meaning from the work differential, or work 1-form,
\begin{equation}
    \omega
    =
    f_x {\mathrm{d}} x + f_y {\mathrm{d}} y + f_z {\mathrm{d}} z
    \label{forcedef}
\end{equation}
which is the integrand for evaluating work along any chosen
path.  The 1-form $\omega$ is said to be exact if Eq.~(\ref{forcedef}) 
equals the differential of some potential function, $\omega= -{\mathrm{d}} \Phi$;
otherwise $\omega$ is inexact.  For our purposes an exact 1-form is the same thing
as a conservative force.

In Euclidean space
one can reinterpret $f_x$, $f_y$, and $f_z$ as components of a vector.
But as a 1-form field, Eq.~(\ref{forcedef}) is constrained by Darboux's
theorem \cite{Edelen,Burke}, which states that every
1-form field $\omega$ can be reduced by some choice of general coordinates
($q_1,q_2,q_3, \dots$) to a shortest canonical form 
$\omega=\omega^{(k)}$ belonging to the
sequence 
\begin{eqnarray}
    \omega^{(1)} 
     &=&
     {\mathrm{d}} q_1 
     ,
     \label{Darboux-one}
     \\
     \omega^{(2)}
     &=&
     q_1 {\mathrm{d}} q_2 
     ,
     \label{Darboux-two}
     \\
     \omega^{(3)}
     &=&
     {\mathrm{d}} q_1+ q_2 {\mathrm{d}} q_3
     \label{Darboux-three}
\end{eqnarray}
In more than three dimensions the list continues with $\omega^{(4)} = q_1
{\mathrm{d}} q_2+ q_3 {\mathrm{d}} q_4$, and so on.

The case $\omega=\omega^{(1)}$, i.e.\ $\omega={\mathrm{d}} q_1$, is exact,
i.e.\ conservative. Not only is $q_1$ a coordinate, but $-q_1$ is the potential
function for $\omega$.  The case $\omega=\omega^{(2)}$, or
$\omega=q_1{\mathrm{d}} q_2$, is reducible to the exact case by an integrating
factor $(1/q_1)$, since $(1/q_1)\omega = {\mathrm{d}} q_2$.  In two dimensions
this exhausts our list, showing that all 1-forms (i.e., differentials) in 2D
are either exact or integrable, a well-known and useful fact in thermodynamics.

In 3D, exact (conservative) and integrable force fields of types
$\omega=\omega^{(1)}$, and $\omega=\omega^{(2)}$ can still occur, but the
most general possibility is the nonintegrable 1-form $\omega^{(3)}$, i.e.\
$\omega= {\mathrm{d}} q_1+ q_2 {\mathrm{d}} q_3$.  The form of a generic force
field is thus more restricted than Eq.~(\ref{forcedef}) would suggest.  In
the case of optical trapping, the optical force field includes a part derivable
from a potential, which we use for our first coordinate: $q_1=-\Phi$.  Following
Eq (\ref{Darboux-three}), we write
\begin{equation}
    \omega
    =
    f{\mathrm{d}} \theta -{\mathrm{d}} \Phi
    \label{threedim}
\end{equation}
so that $q_3=\theta$ is a coordinate which we will below take to be an angle,
and $q_2=f$, the angular force conjugate to $\theta$.  The Frobenius theorem
\cite{Edelen,Burke,Frankel} states that in terms of the exterior product
$\wedge$ and exterior derivative ${\mathrm{d}} $, the work 1-form will be of the
nonintegrable form of Eq.~(\ref{threedim}) whenever $\omega \wedge {\mathrm{d}} \omega
= -{\mathrm{d}} \Phi \wedge {\mathrm{d}} f \wedge {\mathrm{d}} \theta \ne  0$.

In the neighborhood of an equilibrium point, where all components of a force
field vanish, it is natural to interpret $\theta$ in Eq.~(\ref{threedim}) as
an angular variable, motivated by the reduction $-y{\mathrm{d}} x + x{\mathrm{d}} y
=\rho^2{\mathrm d}\theta$.  Nonconservative forces near equilibrium are
circulations aligned with some plane that contains the equilibrium point.
The nonconservative portion of the force can be taken as strictly circular
near the equilibrium point in any coordinates, since a noncircular 2D force
pattern in $x$ and $y$ can be written as
\begin{equation}
    -ay \,{\mathrm d}x  
    +bx \,{\mathrm d}y  
    =
    \tfrac{1}{2}
    (a+b)
    \rho^2 \,{\mathrm d}\theta
    -
    \tfrac{1}{2}
    (a-b)
    \,{\mathrm d}(xy)
    \label{makecircle}
\end{equation}
and the last term can be absorbed into $-d\Phi$. Thus the work 1-form
(\ref{makecircle}) is actually of the type (\ref{threedim}), with $\theta$ a
circular angle in terms of $x$ and $y$.

Thus, to study a nonconservative 3D force about an equilibrium
point, it is sufficient to consider a circular pattern of 
force in some 2D plane containing the equilibrium point.

\begin{figure}
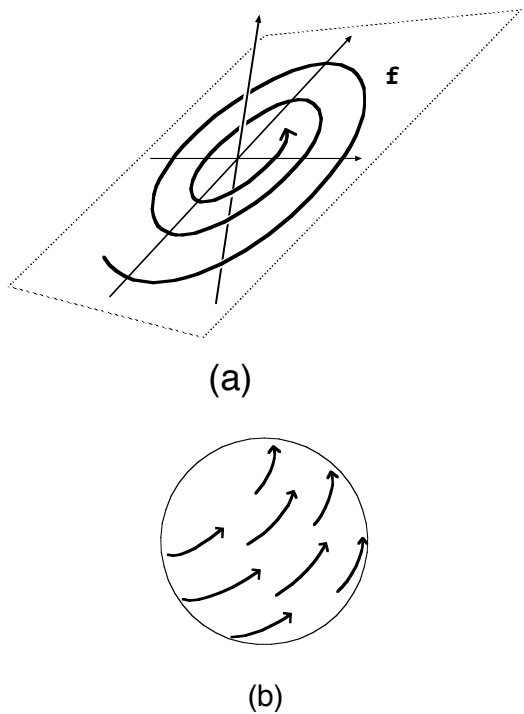

\begin{centering}
\includegraphics[width=0.8\columnwidth]
{FIG01a.pdf}

\vspace{2.0ex}

\includegraphics[width=0.35\columnwidth]
{FIG01b.pdf}

\end{centering}
\caption{
A nonconservative 3D force field near a stable equilibrium point. The circulation of
force may be viewed as purely circular in some plane.  (a) Vector flow-field
point of view. (b) Combed-hair-on-a-sphere point of view.
}
\end{figure}

From another point of view, this picture is consistent with equilibrium-point
analysis of a vector field $\mathbf f$, viewed as flow towards a fixed point
\cite{Jordan} (Fig.~1).  Briefly, with coordinates $q_i$ we have $f_i \approx
M_{ij}\delta q_j$ near the fixed point, defining some real matrix ${\mathsf
M}$.  The eigenvalues of ${\mathsf M}$ are roots of a cubic equation with
real coefficients, and generically yield one real and two complex conjugate
roots whose real parts, together with their eigendirections, correspond here
to a conservative force field.  The imaginary part of the complex pair of
eigenvalues defines circulation in some plane.  The eigendirections of ${\mathsf
M}$ may be stretched and skewed, corresponding to the non-orthogonal coordinates
in (\ref{Darboux-three}).

From yet another point of view, any vector field restricted to
the surface of a sphere (about the equilibrium point) will 
circulate about the sphere in a 2D fashion, according to the
familiar ``combed hair on a sphere'' theorem of mathematics
\cite{Frankel} (Fig.~1).

\section{Nonconservative trapped-sphere model}

For a ``toy model'' realization of a circular nonconservative force we consider
an optically trapped sphere much larger than than the wavelength of light,
so that geometrical optics applies.  In our simplified model we assume no
reflections (perhaps due to a graduated-index boundary) and we trap the sphere
with only a few rays passing almost centrally through the sphere, so that
refraction and trapping force can be calculated in a paraxial-ray approximation.
Our model is certainly contrived, but is very simple to analyze.  In more
realistic situations, whenever there is chirality (handedness) of a system
of light rays near an equilibrium point, we still expect some degree of
circular nonconservative forcing to occur.

\begin{figure}
\begin{centering}
\includegraphics[width=0.7\columnwidth]
{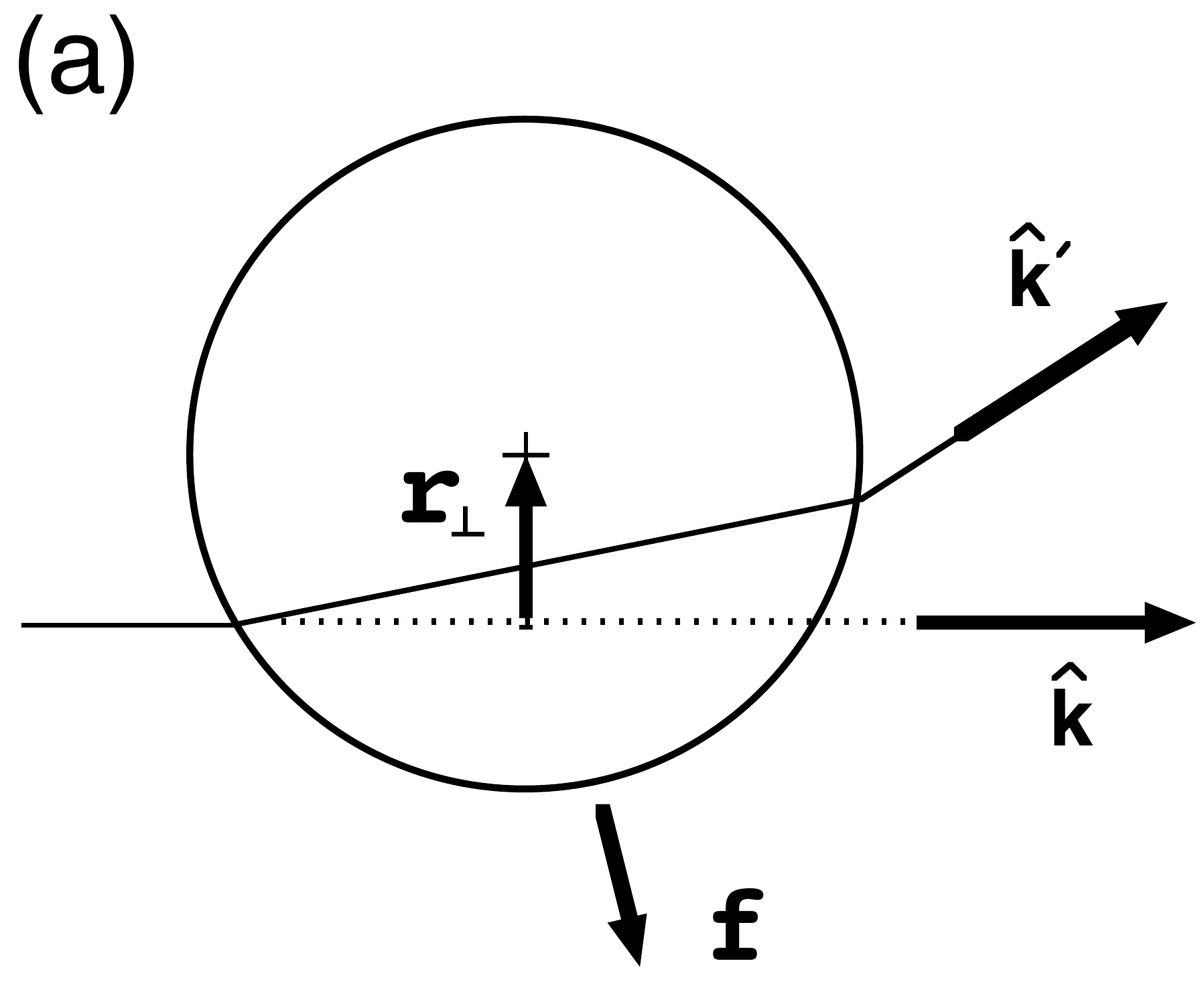}

\vspace{1.0ex}

\includegraphics[width=0.6\columnwidth]
{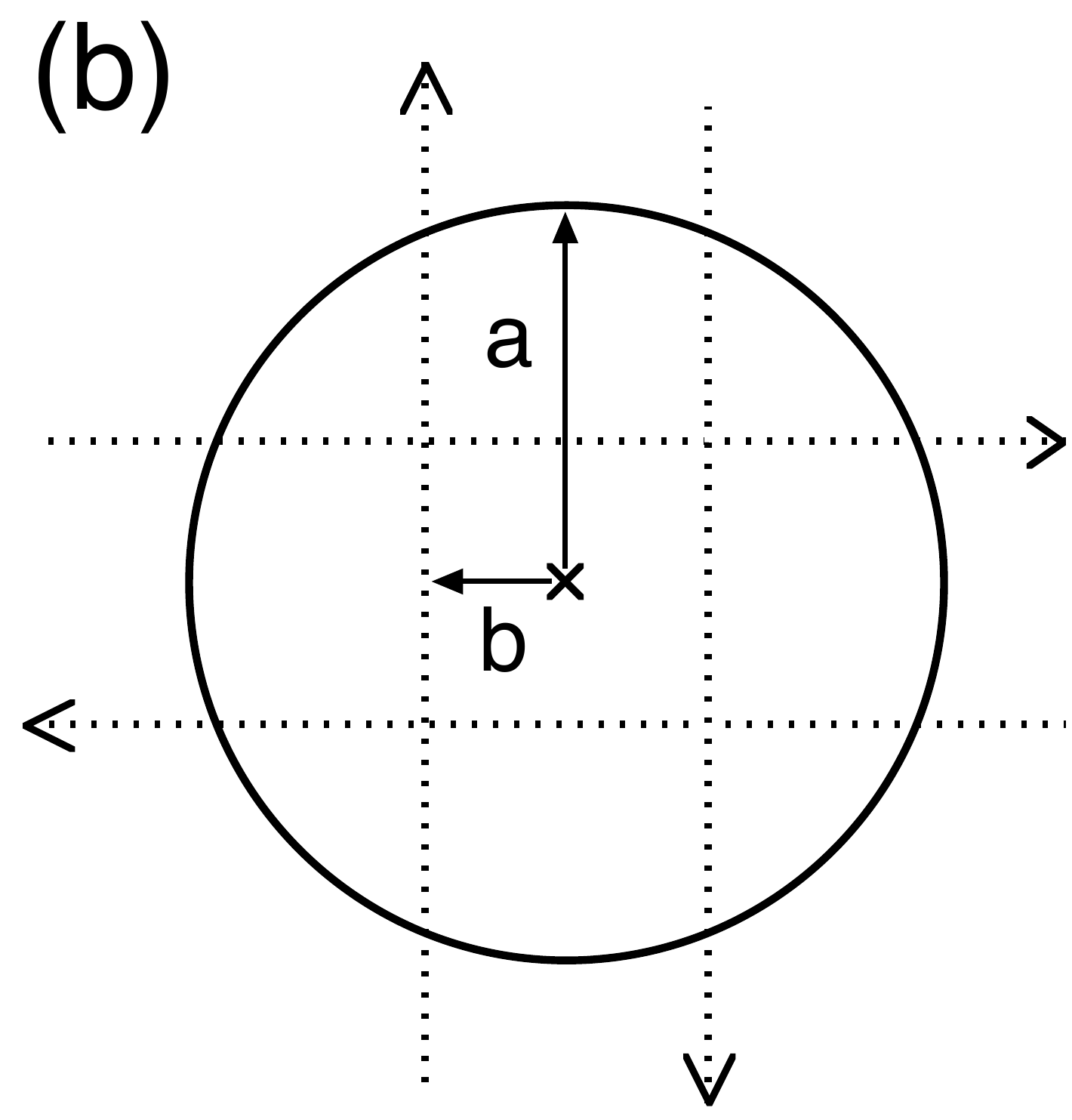}

\end{centering}

\caption{
(a) Refraction of a single ray by a sphere. In the paraxial
limit, ${\mathbf r}_\perp$ is much smaller than the
radius of the sphere.  The force $\mathbf f$ acts to return the sphere center
to the line of $\mathbf{\hat k}$, but also
has a component along $\mathbf{\hat k}$.
(b)~System of four rays for nonconservative trapping.
The limit $b\ll a$ allows a paraxial approximation for each ray.
}
\end{figure}

We need only consider rays within the plane of Fig.~2(a), 
containing the sphere center.  In the figure the displacement of the
sphere center from the ray is exaggerated.  The force of the refracted ray
on the sphere due to momentum transfer is
\begin{equation}
    \mathbf f
    =
    -\frac{I_0}{c} ( \mathbf{\hat k}'-\mathbf{\hat k})
    ,
\end{equation}
where $I_0$ is the power of the ray and $\mathbf{\hat k}$ and $\mathbf{\hat
k}'$ are unit vectors for the original and refracted rays.  If $\Delta\theta$
is the angle between $\mathbf{\hat k}$ and $\mathbf{\hat k}'$, then
\begin{equation}
    \mathbf f
    =
    -\frac{I_0}{c}
    \big[(\cos\Delta\theta-1)\,\mathbf{\hat k}
    +(\sin\Delta\theta)\,\mathbf{\hat r}_\perp\big]
    \label{force}
\end{equation}
where $\mathbf{\hat r}_\perp$ is the direction of $\mathbf r_\perp$ (see Fig.~2(a)).
We set $\sin\Delta\theta\approx\Delta\theta$ and $\cos\Delta\theta-1 \approx
-\Delta\theta^2/2$, and use paraxial ray transfer matrices \cite{Burch} to
express $\Delta\theta$ and $\mathbf f$ in terms of the displacement $\mathbf
r$ of the sphere.  These matrices act on a vector $(\Delta\theta,h)$ whose
components are the angle with and displacement from the optical axis.  If
matrices ${\mathsf F}_{\text{in}}$ and ${\mathsf F}_{\text{out}}$ describe
refraction at the convex surfaces of our sphere of radius $a$, and and matrix
${\mathsf P}_{2a}$ describes paraxial propagation by a distance $2a$, the ray
matrix for the sphere is
\begin{equation}
    {\mathsf F}_{\text{out}}{\mathsf P}_{2a}{\mathsf F}_{\text{in}}
    =\frac{2}{n}\begin{bmatrix}
    1-n/2 & -(n-1)/a \\
    a & 1-n/2 \end{bmatrix}
\end{equation}
with $n$ the relative refractive index of the sphere.
In our case, $h=-r_\perp$ gives
\begin{equation}
    \Delta\theta
    =
    \frac{2(n-1)}{na}\,r_\perp
\end{equation}
for the angular deviation.
For simplicity we set $n=2$ to find
\begin{equation}
    \mathbf f\approx 
    \frac{I_0}{c}
    \left[ \,
    \frac{r_\perp^2}{2a^2} \,\mathbf{\hat k} - \frac{\mathbf r_\perp}{a} 
    \right]. 
    \label{rayforce}
\end{equation}
Eq.~(\ref{rayforce}) resolves $\mathbf f$ into a conservative restoring force,
proportional to $-\mathbf r_\perp$, and a nonconservative quadratic force,
pushing the sphere along the ray direction $\mathbf{\hat k}$ whenever the
sphere center is displaced off the ray axis.

The system of four rays shown in Fig.~2(b) results in a purely
circular nonconservative force on the sphere. Despite the
clockwise circulation of rays, there is no torque on the sphere
within this ray-optical model: it is the nonconservative net force on
the sphere that we study here.  We separate oppositely
directed ray pairs by a distance $2b$, exaggerated in the figure,
with $b\ll a$ so that our paraxial analysis remains valid.  The
ray in the $+x$ direction is described by
\begin{equation}
    \mathbf{\hat k} = \mathbf{\hat x},
    \quad
    \mathbf r_\perp = (y-b) \, \mathbf{\hat y} + z\mathbf{\hat z}
\end{equation}
and similarly for the other rays.  Summing the forces from
Eq.~(\ref{rayforce}), the total force on the sphere is
\begin{equation}
    \mathbf f 
    =  
    \frac{2I_0}{c a} 
    \left[ 
    \frac{b}{a}
    (x\mathbf{\hat y}-y\mathbf{\hat x})-(x\mathbf{\hat x}+ y\mathbf{\hat y} + 2z\mathbf{\hat z}) 
    \right]
    .
\end{equation}
The work differential, or work 1-form, is
\begin{equation}
    \omega 
    =
    \frac{2bI_0}{c a^2} 
    \, \rho \, {\mathrm{d}}\theta
    -
    {\mathrm{d}} \! \left[ 
    \frac{I_0}{c a} 
    (x^2 +y^2 +2z^2)
    \right] 
\end{equation}
which is of the form of Eq.~(\ref{threedim}), $\omega = f{\mathrm{d}}\theta -{\mathrm{d}}\Phi$.

\section{Nonconservative motion of a diffusing particle}

To look for spectral features of trapping in a nonconservative field, we
consider diffusion through a fluid of a particle with drag coefficient $\gamma$.
At temperature $T$, the diffusion coefficient is $\gamma/k_{\scriptscriptstyle
\mathrm B}T$, an Einstein relation \cite{Nelson}.  Near an equilibrium point
we consider 2D motion ${\mathbf r}(t)$ in a plane of circulation where the
nonconservative component of force is
\begin{equation}
    \mathbf f 
    =
    \xi \mathbf{\hat z}\!\times\! \mathbf r
    =
    \xi \,\rho\, \mathbf{\hat \theta}
\end{equation}
where $\xi$, the strength of the circular forcing, has units of force per
distance.  As a 1-form, 
\begin{equation}
    \omega 
    = \xi\,\rho \,{\mathrm{d}}\theta
    .
\end{equation}
First we
check that there is no pathology regarding work done by the nonconservative
force at small length scales, where a diffusing particle executes spatial
cycles with a diverging frequency.  For this, imagine the particle constrained
to move on a circle of radius $\epsilon$, like a bead sliding on a circular
wire.  In the absence of the external force, the time to diffuse around the
circle is $\delta t\sim (\gamma/k_{\scriptscriptstyle\mathrm B}T)\,\epsilon^2$.
In the presence of a nonconservative force, the rates
of positive and negative circling are enhanced and suppressed according to
\begin{equation}
    \nu_\pm
    \sim
    \frac{k_{\scriptscriptstyle \mathrm B}T}{\gamma\epsilon^2}
    \, \exp
    \Big(
    \! \pm\frac{\Delta W}{k_{\scriptscriptstyle \mathrm B}T}
    \Big)
\end{equation}
where $\Delta W$ is the work done by the nonconservative force in a ``$+$''
cycle; for our force this is $\Delta W = 2\pi \epsilon f \sim \xi\epsilon^2$.
The net ``$+$'' rate is $\Delta\nu = \nu_+ \!- \nu_- \sim \Delta
W/\gamma\epsilon^2$, implying that the dissipated power is
\begin{equation}
    \Delta\nu
    \,\Delta W
    \sim
    \frac{\xi^2 \epsilon^2}{\gamma}
    .
\end{equation}
We conclude that the power supplied by the nonconservative force vanishes at
small scales, as the square of the spatial scale of the motion.

We turn to the spectral properties of the  diffusing particle, and first
establish formalism \cite{Gittes-noise} with force and motion only in the $x$
direction.  For a force $f(t)$ and motion $x(t)$ that extend over all time,
the spectral density $(fx)_{\omega}$ is defined by \cite{Landau}
\begin{equation}
    \langle {f}_{\omega}x^*_{\omega'}\rangle
    =
    2\pi \, ({f}x)_{\omega}\,\delta(\omega-\omega')
\end{equation}
where $f_{\omega}$ and $x_{\omega}$ are Fourier transforms of functions
truncated outside of a time $T$.  Throughout, $(\dots)_\omega$ specifies a
spectral density, whereas $\langle\dots\rangle$ is an average.  Standard
manipulations \cite{Landau}, or more simply the formal replacement $\langle
f_{\omega}x^*_{\omega'}\rangle = T\, (fx)_{\omega}$, give the rate of work
done by $f$ as
\begin{equation}
    \Big\langle \frac{dW}{dt} \Big\rangle 
    \;=\; 
    \frac{1}{2\pi} 
    \int
    (fv)_\omega
     \,
    {\mathrm{d}}\omega
    \;=\;
    \frac{1}{2\pi} \int 
    i\omega (fx)_\omega
    \,
    {\mathrm{d}}\omega
    .
\end{equation}
Spectral densities integrate to correlations or mean squared fluctuations
\cite{Landau} In the 1D case, if a thermal Nyquist force $N(t)$ acts on the
object, with $(N^2)_\omega = 2\gamma k_{\scriptscriptstyle\mathrm B}T$ (with
$\gamma$ the drag coefficient) then in an external conservative force $f =
-\kappa_x x$, with $\alpha_x=\kappa_x/\gamma$, the equation of motion $N =
\kappa_x x +  \gamma \dot x$ gives us
\begin{eqnarray}
    N_\omega
    &=&
    -i\gamma (\omega+i\alpha_x)
    x_\omega
    \label{EOM1}
    \\
    \langle N_\omega N^*_\omega\rangle 
    &=&
    \gamma^2 (\alpha_x^2 +\omega^2)
    \langle x_\omega x^*_\omega\rangle
    \label{EOM1SQ}
\end{eqnarray}
The replacements
$\langle N_{\omega}N^*_{\omega'}\rangle = T\, (N^2)_{\omega}$ and
$\langle x_{\omega}x^*_{\omega'}\rangle = T\, (x^2)_{\omega}$
immediately lead to
\begin{equation}
    (x^2)_\omega
    =
     \, \frac{2k_{\scriptscriptstyle\mathrm B}T/\gamma}{\alpha_x^2 +\omega^2}
     .
     \label{1DLorentzian}
\end{equation}
This Lorentzian spectrum integrates to 
\begin{equation}
    \langle {x^2} \rangle 
    = 
    \frac{1}{2\pi} 
    \int
    (x^2)_\omega
     \,
    {\mathrm{d}}\omega
    \;=\;
    \frac{k_{\scriptscriptstyle\mathrm B}T}{\kappa_x}
     \label{xx1D}
\end{equation}
as it should by the equipartition theorem \cite{Gittes-noise}.  Since
$(fx)_\omega = -\kappa_x (x^2)_\omega$ is an even function of $\omega$, the
rate of external work vanishes.

Now include a circular external force. With an
isotropic conservative force
$-\kappa \mathbf r$,
the vector equation of motion
    ${\mathbf N} 
    =
    \kappa {\mathbf r} 
    - \xi\mathbf{\hat z}\times{\mathbf r}
    +\gamma \dot{\mathbf r}$
gives us
\begin{equation}
    {\mathbf N}_\omega 
    =
    (\alpha-i\omega  - \eta \,\hat{\mathbf z}\times)
    \, {\mathbf r}_\omega
    \label{EOMC}
\end{equation}
with ``$\times$'' a cross product, and where $\alpha=\kappa/\gamma$, and $\eta
= \xi/\gamma$.  In the absence of thermal forces, the particle spirals to the
origin with angular frequency $\eta$ and a radius proportional to $\exp(-\alpha
t)$, similar to Fig.\ 1(a).  Using the column vectors
\begin{equation}
    {\mathsf N}_\omega
    = 
    \bigg[\!
    \begin{array}{c}
        N_{x,\omega} \\
        N_{y,\omega} 
    \end{array}
    \!\bigg]
    ,\quad
    {\mathsf r}_\omega
    =
    \bigg[\!
    \begin{array}{c}
        x_\omega \\
        y_\omega 
    \end{array}
    \!\bigg]
\end{equation}
we can analyze an anisotropic confining potential, with Eq.~(\ref{EOM1})
generalized to
\begin{equation}
    {\mathsf N}_\omega 
    = 
    -i\gamma{\mathsf M}{\mathsf r}_\omega
    ,\quad
    {\mathsf M}
    =
    \bigg[\!
    \begin{array}{cc}
        \omega\!+\!i\alpha_x & i\eta  \\
        -i\eta & \omega\!+\!i\alpha_y  
    \end{array}
    \!\bigg]
    \label{EOM2}
\end{equation}
and $\alpha_y = \kappa_y/\gamma$. 
To find the power spectrum we form the averaged product
\begin{equation}
    \langle
    {\mathsf N}_\omega {\mathsf N}^\dagger_\omega 
    \rangle
    =\;
    \gamma^2
    \,
   {\mathsf M} 
    \langle
   {\mathsf r}_\omega {\mathsf r}^\dagger_\omega
    \rangle
   {\mathsf M}^\dagger
   ,
   \label{7}
\end{equation}
a generalization of Eq.~(\ref{EOM1SQ}).
In terms of spectral densities we have
\begin{equation}
    2\gamma k_{\scriptscriptstyle\mathrm B}T
    \,{\mathsf 1}
    \;=\;
    \gamma^2
    \,
   {\mathsf M} 
   ({\mathsf r} {\mathsf r}^\dagger)_\omega
   {\mathsf M}^\dagger
   \label{matrixeqn}
\end{equation}
where ${\mathsf 1}$ is the identity matrix,
and the spectral density matrix is
\begin{equation}
   ({\mathsf r} {\mathsf r}^\dagger)_\omega
   =
    \bigg[\!
    \begin{array}{cc}
        (x^2)_\omega & (xy)_\omega \\
        (yx)_\omega & (y^2)_\omega
    \end{array}
    \!\bigg]
    .
   \label{powermatrix}
\end{equation}
Eq.~(\ref{matrixeqn}) is easily solved for $({\mathsf r} {\mathsf r}^\dagger)_\omega$,
giving diagonal elements
\begin{equation}
    (x^2)_\omega
    =
    \frac{2k_{\scriptscriptstyle\mathrm B}T}{\gamma}
    \frac{
        \omega^2\!+\!\alpha_y^2\!+\!\eta^2 
    }
    {
    [\,
    \omega^4
    \!+\!
    (\alpha_x^2 \!+\! \alpha_y^2 \!-\! 2\eta^2)\omega^2
    \!+\!
    (\eta^2\!+\!\alpha_x\alpha_y)^2
    \,]
    }
    \label{diagonal}
\end{equation}
which generalizes Eq.~(\ref{1DLorentzian}); $(y^2)_\omega$ is the same with
$\alpha_x$ and $\alpha_y$ interchanged.

Eq.~(\ref{diagonal}) represents the power spectrum of $x$-motion of a trapped
particle with a nonconservative force ($\eta\ne 0$) and 2D anisotropy
($\alpha_x\ne\alpha_y$) both in the $xy$ plane.  Nonconservative forces make
this power spectrum non-Lorentzian even in an isotropic trap.  Other interesting
effects such as inertial hydrodynamics and material properties
\cite{Flyvbjerg,Gittes-rheo} will of course also invalidate a simple Lorentzian
spectrum. Carrying out the $\omega$ integral in Eq.~(\ref{xx1D}), we obtain
\begin{equation}
    \langle x^2 \rangle
    \;=\;
    \frac{k_{\scriptscriptstyle\mathrm B}T}{\kappa_x \!+\! \kappa_y}
    \bigg[
    1 + \frac{\xi^2 \!+\kappa_y^2}{\xi^2\!+\!\kappa_x\kappa_y}
    \bigg]
    .
    \label{xx}
\end{equation}
For $\langle y^2 \rangle$, $\kappa_x$ and $\kappa_y$ are interchanged.
In the isotropic case, $\kappa_x \!=\!  \kappa_y=\! \kappa$, the nonconservative
circulation preserves $\langle x^2 \rangle = k_{\scriptscriptstyle\mathrm
B}T/\kappa$.  For large values of $\xi$,
\begin{equation}
    \langle x^2 \rangle = 
    \frac{2k_{\scriptscriptstyle\mathrm B}T}{\kappa_x \!+\! \kappa_y}
    , 
    \quad
    \xi \gg \kappa_x,\kappa_y
    \label{circlimit}
\end{equation}
and the
potential is effectively being averaged by rapid circulation.  However, for
all values of $\xi$, $\kappa_x$, and $\kappa_y$ we find
\begin{equation}
    \big\langle \tfrac{1}{2}\kappa_x x^2 
    +
    \tfrac{1}{2}\kappa_y y^2 
    \big\rangle
    \;=\;
    k_{\scriptscriptstyle\mathrm B}T
\end{equation}
so that the equipartition of potential energy is unaffected by nonconservative
circulation, even though the nonconservative flows do not follow equipotentials
in an anisotropic trap.  

Details of the flow pattern follow from the off-diagonal element of $({\mathsf
r} {\mathsf r}^\dagger)_\omega$,
\begin{equation}
    (xy)_\omega
    =
    \frac{2k_{\scriptscriptstyle\mathrm B}T}{\gamma}
    \frac{
      \eta(\alpha_y\!-\!\alpha_x) \!-\! 2i\eta\omega 
    }
    {[
    \omega^4
    \!+\!
    (\alpha_x^2 \!+\! \alpha_y^2 \!-\! 2\eta^2)\omega^2
    \!+\!
    (\eta^2\!+\!\alpha_x\alpha_y)^2
    ]}
    .
    \label{offdiagonal}
\end{equation}
obtained by solving Eq.~(\ref{matrixeqn}).  Integrating this over $\omega$ yields
\begin{equation}
    \langle xy \rangle
    \;=\;
    \frac{k_{\scriptscriptstyle\mathrm B}T}{\kappa_x \!+\! \kappa_y}
    \;\frac{\xi(\kappa_y\!-\!\kappa_x)}{\xi^2\!+\!\kappa_x\kappa_y}
    \label{xy}
\end{equation}
Using this result we can generalize Eq.~(\ref{xx}) to mean squared radius
at any 2D angle.
For Gaussian statistics such as we have here,
the probability distribution
\begin{eqnarray}
    P(x,y) 
    &\propto&
    \exp[-\tfrac{1}{2} 
    {\mathsf r}^{\scriptscriptstyle{\top}}{\mathsf G}\,{\mathsf r}]
    \label{binormal}
\end{eqnarray}
yields a matrix of averages
$\langle {\mathsf r}{\mathsf r}^{\scriptscriptstyle{\top}} \rangle
= {\mathsf G}^{\scriptscriptstyle{-1}}$,
and a unit ellipse defined by
\begin{equation}
    {\mathsf r}^{\scriptscriptstyle{\top}}{\mathsf G}\,{\mathsf r} 
    \;=\;
    {\mathsf r}^{\scriptscriptstyle{\top}}
    \langle {\mathsf r}{\mathsf r}^{\scriptscriptstyle{\top}} \rangle^{-1}
    \,{\mathsf r} 
    \;=\;
    1
    .
\end{equation}
As a function of
angular direction $\theta$,
the squared radius of the ellipse is
\begin{equation}
    r^2
    \;=\;
    \frac{
    \langle x^2 \rangle
    \langle y^2 \rangle
    -
    \langle xy \rangle^2
    }{
    \langle y^2 \rangle \cos^2\theta 
    - 2\langle xy \rangle \cos\theta \sin\theta 
    + \langle x^2 \rangle \sin^2\theta
    }
    \label{rho2theta}
\end{equation}
into which $\langle x^2 \rangle$, $\langle y^2 \rangle$, and $\langle xy
\rangle$ can be inserted from Eqs.~(\ref{xx}) and (\ref{xy}).  The resulting
ellipses are plotted in Fig.~3 for various values of $\xi$.  The r.m.s.\
radius drawn in the figure is necessarily a line of flow.  For increasing
$\xi$ the flow pattern does not follow the $\xi=0$ equipotential lines, but
tilts by $\theta_\xi$ in the direction of the circular forcing, with
\begin{equation}
    \tan 2\theta_\xi
    \;=\;
    \frac{2\xi}{(\kappa_x+\kappa_y)}
    .
\end{equation}
The flow pattern also decreases in eccentricity with $\xi$ until, at high
circular forcing, we recover the circularly symmetric distribution described
by Eq.~\ref{circlimit}.

\begin{figure}
\begin{centering}

\includegraphics[width=0.9\columnwidth]
{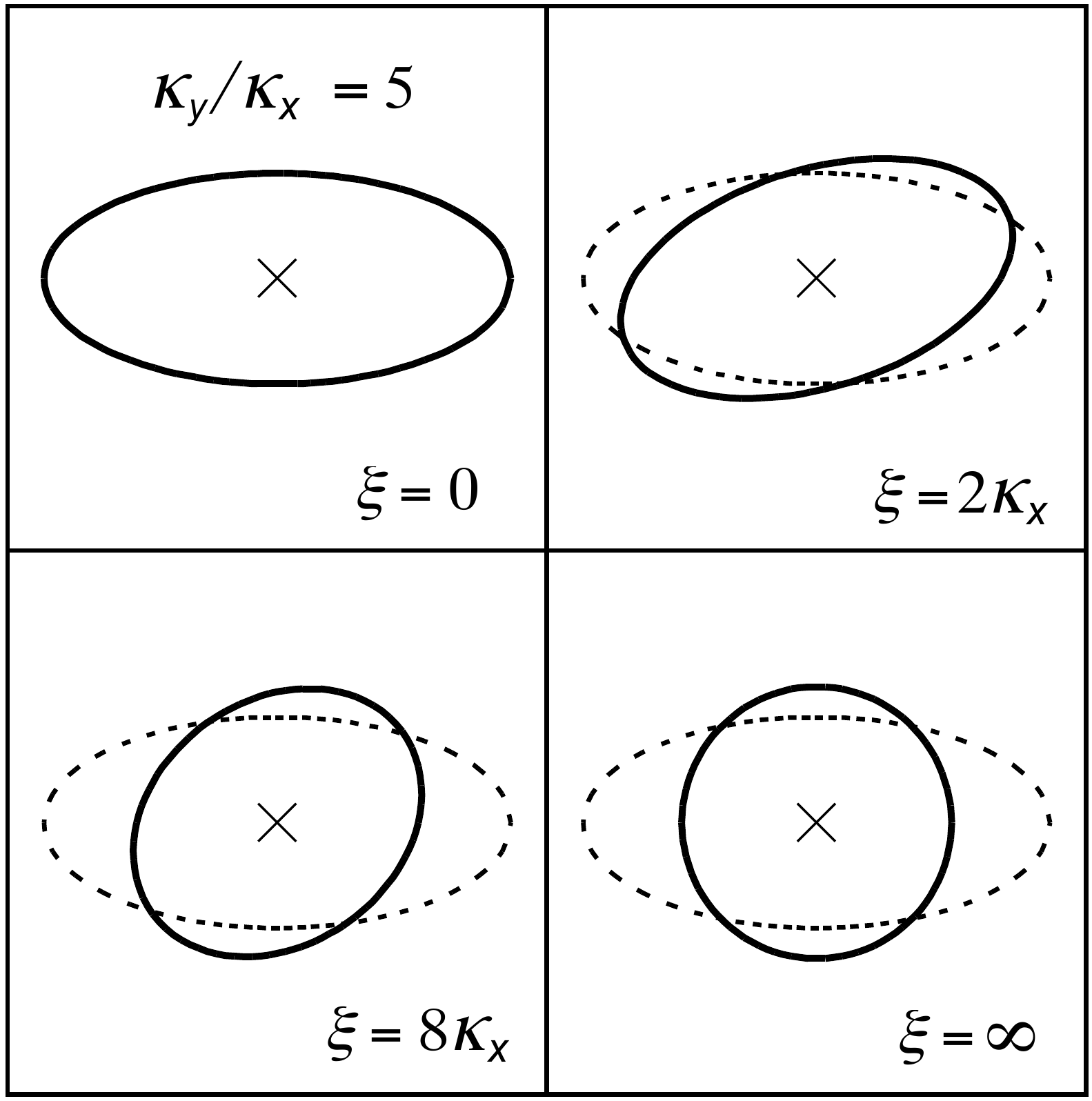}

\end{centering}

\caption{
R.m.s.\ radius $\langle r^2\rangle^{1/2}$ (solid curves) of the position
distribution $P(x,y)$ of a diffusing particle in a trap with stiffness
anisotropy $\kappa_y/\kappa_x = 5$, for various values $\xi$ of counterclockwise
circular forcing in the $xy$ plane.  The solid curves are lines of flow which,
for $\xi\ne 0$, do not follow the $\xi=0$ equipotential contour (dotted
curves).  Circular flow at $\xi=\infty$ (final panel, solid curve) has the
r.m.s.\ radius \mbox{$[2k_{\scriptscriptstyle\mathrm B}T/(\kappa_x \!+\!
\kappa_y)]^{1/2}$}.
}
\end{figure}

The rate at which work is done on the system is
\begin{equation}
    \Big\langle \frac{dW}{dt} \Big\rangle 
    \;=\; 
    \frac{1}{2\pi} 
    \int
    ({\mathbf f}\!\cdot{\mathbf v})_\omega
     \,
    {\mathrm{d}}\omega
    \label{extwork}
\end{equation}
where $({\mathbf f}\!\cdot{\mathbf v})_\omega$ is a spectral density.
In terms of Fourier transforms,
\begin{eqnarray}
    {\mathbf f}_\omega
    \cdot {\mathbf v}^*_\omega
    &=&
    i\omega\,\xi
    \,
    \bigg[\!
    \begin{array}{c}
        x_\omega \\
        y_\omega 
    \end{array}
    \!\bigg]^\top
    \bigg[\!
    \begin{array}{cc}
        0 & 1  \\
        -1 & 0 
    \end{array}
    \!\bigg]
    \bigg[\!
    \begin{array}{c}
        x^*_\omega \\
        y^*_\omega 
    \end{array}
    \!\bigg]
    \nonumber
    \\
    &=&
    i\omega\xi( x_\omega y^*_\omega - y_\omega x^*_\omega )
    \nonumber
    \\
    &=&
    -2\xi\omega\,{\mathsf{Im}}\,x_\omega y^*_\omega
    \label{workrate}
\end{eqnarray}
from which we infer the spectral density
\begin{equation}
    ({\mathbf f}\!\cdot{\mathbf v})_\omega
    = -2\xi\omega \; {\mathsf{Im}}\,(xy)_\omega
    .
    \label{fvpower}
\end{equation}
Using Eq.~(\ref{offdiagonal}) for the right-hand side and performing the integral
in Eq.~(\ref{extwork}) gives
\begin{equation}
    \Big\langle \frac{dW}{dt} \Big\rangle 
    \;=\;
    \frac{4k_{\scriptscriptstyle\mathrm B}T \,\xi^2}{\kappa_x \!+\! \kappa_y}
    \,.
\end{equation}
We find that the rate of work dissipated in the system by the circular
nonconservative force is directly proportional to temperature, reflecting the
enhancement of induced drift velocities by thermal spreading from the zero-force
point.  In contrast, nonconservative toroidal circulation due to axial radiation
pressure \cite{Roichman,Sun} leads to a squared dependence on temperature.
Depending on the rate of outward heat flow from the optical focus
\cite{Peterman} the proportionality of dissipation to temperature raises the
possibility of an instability towards runaway heating by a nonconservative
component of the trapping force.

\section{Summary}

We have characterized the simplest 3D nonconservative force field, near a
stable fixed point, as an anisotropic conservative force plus a nonconservative
circular force.  We constructed a simple trapping model exhibiting such
forcing, and presented signatures of particle displacement and nonconservative
dissipation in the presence of nonconservative forcing.  In most optical
trapping experiments, simple circular forcing about a stationary point may
not be easily observable given the more dominant radiation pressure effects
that have been observed by others \cite{Roichman,Sun}. Specifically designing
chiral trapping geometries for particles and minimizing reflections may allow
observation of circular forcing. More generally, however, the geometrical
picture we have constructed and the thermal and spectral results we have
obtained may be useful in other situations, even outside of optics, where
locally nonconservative microscopic forces act.

\section*{Acknowledgements}
The authors would like to thank their colleagues in the Department of Physics 
and Astronomy at Washington State University for support and helpful
conversations.

\end{document}